\documentclass[epj]{svjour}
\usepackage{graphics}
\usepackage{graphicx}
\usepackage{epsf}
\usepackage{bm}
\usepackage{amssymb}
\usepackage{amsmath}
\begin{document}
\title{Effective field theory for spinor dipolar Bose Einstein condensates}
\author{M. Takahashi\inst{1}, Sankalpa Ghosh\inst{2}, T. Mizushima\inst{1}, and
K. Machida\inst{1}
}
% etc
%}                     % Do not remove
\institute{Department of Physics, Okayama University, Okayama 700-8530, Japan
\and Department of Physics,
Indian Institute of Technology, Delhi,
Hauz Khas,
New Delhi 110016,
India}
\date{Received: date / Revised version: date}
%\begin{document}
\def \beq{\begin{equation}}
\def \eeq{\end{equation}}
\def \bea{\begin{eqnarray}}
\def \eea{\end{eqnarray}}
\def \bs{\boldsymbol}
%\title{Effective field theory for spinor dipolar Bose Einstein condensates}
%
%\author{M. Takahashi$^1$, Sankalpa Ghosh$^2$, T. Mizushima$^1$, and
%K. Machida$^1$}
%
%\affiliation{$^1$Department of Physics, Okayama University,
%Okayama 700-8530, Japan}
%\affiliation{$^2$Department of Physics,
%Indian Institute of Technology, Delhi,
%Hauz Khas,
%New Delhi 110016,
%India}
\abstract{
We show that the effective theory of long wavelength low energy behavior of a dipolar Bose-Einstein condensate(BEC) with
large dipole moments (treated as a classical spin) can be modeled using  an extended  Non-linear sigma model (NLSM) like energy functional with an additional non-local term that represents  long ranged anisotropic dipole-dipole interaction. Minimizing this effective energy functional we calculate the density and spin-profile of the dipolar Bose-Einstein condensate in the mean-field regime for various trapping geometries. The resulting configurations show strong intertwining between the spin and mass density of the condensate, transfer between spin and orbital angular momentum in the form of Einstein-de Hass effect, and novel topological properties. We have also described the theoretical framework
in which the collective excitations around these mean field solutions can be studied and discuss some examples qualitatively. }
%\PACS{
%      {PACS-key}{describing text of that key}   \and
%      {PACS-key}{discribing text of that key}
%     } % end of PACS codes
%\pacs{03.75.Mn, 03.75.Hh, 67.57.Fg}
\maketitle
\section{INTRODUCTION}
$O(3)$ Non-linear sigma model (NLSM) is one of the most studied exactly solvable model in the non-linear field theory. Particularly in two spatial dimension all its finite energy, stable topologically non-trivial solutions are analytically known due to the pioneering work by Belavin and Polyakov \cite{nlsm,doug}. In more than two dimensions no such finite energy stable topologically non- trivial solution exists according to a theorem by Derrick and Hobart. However,
in spinorial Bose-Einstein condensates, for the relevant time scale for experimental systems, non-trivial texturing can occur in the order parameter field also in three spatial dimensionsn \cite{3dskyrmion} and other type of interaction may actually stabilize them.
In real physical systems, effective field theories describing a number of low dimensional condensed matter systems, such as single and multiple layer quantum Hall systems \cite{Sondhi,Moon,Ghosh}, spinorial BECs \cite{ohmi,ho}etc.,  under various approximations often turn out to be various extensions of NLSM. Of interest to us in this present paper are spinorial BECs, in which all hyperfine states are liberated and  the order parameter of the condensates is described by a multi-component vector. This is a fertile ground of the manifestation of the various topological phases of NLSM \cite{ohmi,ho}. Even though some of these configurations \cite{3dskyrmion} are actually metastable, their study reveals interesting dynamics related to the internal structures of such spinorial condensates.
Spinorial BECs have been realized in fully optically trapped $^{87}$Rb and $^{23}$Na gases with the hyperfine spin $F = 1$ and $F = 2$ \cite{ketterle1,ketterle2} respectively. In all these experiments the dominant interatomic interaction is however
short range interactions and is characterized by the corresponding  $s$-wave
scattering lengths in various spin channels. The other type of interactions, most prominent of
which is the dipole-dipole (d-d) interaction among atoms
can generally be neglected under most of the current experimental situations.
This situation however changed  since
Bose-Einstein condensates that exhibits such dipole-dipole interaction  has recently been achieved by  Griesmaier {\it et al.} by cooling  $^{52}$Cr atom gases in a magnetic trap, whose magnetic moment per atom is $6 \mu_B$ (Bohr magneton) \cite{griesmaier}.
%with the observation of Bose-Einstein condensation in  $^{52}$Cr atom  \cite{griesmaier} with %hyperfine spin $F=3$ and having considerable dipole-dipole interaction which is long-ranged and %anisotropic. In this article we have taken this analogy further by showing that the long-wavelength %behavior of such spinorial BEC with large spin and significant dipole-dipole interaction can be %represented by an extension of NLSM which includes an additional non-local
% term representing the anisotropic long ranged dipole-dipole interaction. The effect of such terms %on the prototype ground state of NLSM is dramatic as we shall see.
These experiments are generally performed in presence of a strong magnetic field where all the dipoles are polarized and completely frozen at the typical experimental temperature. Nevertheless, there has been already emerging several novel aspects associated with larger magnetic moment in $^{52}$Cr atom gases \cite{stuhler,Lahaye}, which is caused by the magnetic dipole-dipole (d-d) interaction. Many theoretical studies for such dipolar BECs that have been done \cite{baranov} are in this limit where the spin degrees of  freedom is frozen and one has to deal with a scalar order parameter.
In contrast, the theoretical studies of spinor BECs with the d-d interaction are rather limited and few papers exist mostly for the $F = 1$ case \cite{kawaguchi,pu,cheng,venga,Lama}. To study the spinor dipolar BEC with arbitrary value of the spin $F$, one has to handle $2 F + 1$ components of spinor
and $F + 1$ interaction channels. Even for $^{52}$Cr case ($F = 3$), we have
to solve $7$ different equations at the same time in the $4$ dimension
interaction parameter space \cite{Santos,diener}. This is quite formidable and complicated. Moreover, interesting topological properties as well as spin-texturing aspects of the resulting ground states configurations are not readily tractable. An alternative approach is to look at the
large spin-limit. From the spin-wave analysis of a magnetic systems we know that
such analysis works very well when the spin fluctuations is essentially a low energy and long-wavelength process as in the case of large spin. Even for small spin sometimes it works
very well. Using this approach, in this paper we model
the spinor dipolar Bose-Einstein  condensates  by treating the spin or the dipole moment of the atom as a freely rotating classical unit vector, namely a classical spin. The resulting theory that represents only the low-energy long-wavelength behavior of the spin fluctuation in such dipolar condensate is an effective energy functional whose leading terms resemble those of Non-linear sigma model (NLSM) hamiltonian embedded in three spatial dimensions, but accompanied by a non-local term that represents the effect of long-range d-d interaction. Since the original NLSM is rotationally invariant this additional term helps us to understand the effect of symmetry breaking anisotropic long-ranged interaction on the rotationally invariant spin-textures in general. We shall particularly point out
some interesting topological features of such solutions, interplay between spin and mass density, interplay between spin and orbital angular momentum. 

Some results have been already reported in a
recent letter \cite{takahashi}. Here we provide a more detailed version of this effective theory and its outcomes by exploring the ground states over a larger region of the parameter space, plot of the resulting ground states and their associated topological properties and how they changed as function of the trap geometry. We  
additionally provide a theoretical framework in which collective excitations can be studied and  a detailed comparison with other methods available to study similar systems.

The rest of the paper is organized as follows. We start with a brief
introduction of the non-linear sigma model and its interesting properties in Sec. \ref{sec:model}.  In the same section we shall the describe the full second quantized hamiltonian  for the spinor-dipolar BECs and the resulting effective energy functional in the
limit of large dipole moments. Then  we shall discuss in Sec. \ref{sec:result} the ground states obtained by minimizing this effective energy functional for various trap geometries. We shall particularly emphasize the ground state for pancake trap and cigar shaped trap, two most widely used geometries. After this Sec. \ref{sec:phases}  will contain the issue of stability of such condensates and the related quantum phase transitions. In the next section ( Sec. \ref{sec:collective}) we provide the theoretical framework in which the collective excitations over these phases can be studied.
This is followed by a comparison with a set of other works that addressed  similar systems (Sec. \ref{sec:compare}). 
We shall conclude these discussions by pointing out the experimental implications of our findings and the possibilities of future studies.
\section{NLSM AND THE EFFECTIVE ENERGY FUNCTIONAL FOR SPINOR DIPOLAR CONDENSATE}\label{sec:model}
We start with a brief description of the prototype non-linear sigma model and properties of its solutions \cite{nlsm}.
The model contains real scalar fields $\phi^{\alpha}(\bf{x})$ with $\alpha=1,2,3$ with the following
non-linear constraint being imposed at each co-ordinate point $\bf{x}$
\beq \phi^{\alpha}(\bf{x})\phi^{\alpha}(\bf{x}) = 1  \label{nlsm1} \eeq
The minimal energy functional that is invariant under global $O(3)$ rotation in the internal space of these scalar fields is  given by
\beq E = \rho_s \int d\bf{x} (\partial_{\mu} \phi^{\alpha})^2 \label{nlsmenergy} \eeq
Here $\rho_s$ is a constant and $\mu$ refers to the co-ordinate variable. The resulting field
equations are non-linear because the above energy functional (\ref{nlsmenergy}) need to be extremized after implementing the normalization constraint (\ref{nlsm1}). Particularly in two-dimension these equations have stable topologically non-trivial solutions (for
a review see ref. \cite{doug}).
Whereas stable solutions are available in $3D$ only in presence of an additional gauge interaction
\cite{Fadeev}. In the dipolar BECs the constant value of the dipole moment of each atom implements the above non-linear constraint (\ref{nlsm1}) and the gradient part of the effective energy functional is manifestly same as the $O(3)$ invariant NLSM model. However the effective energy
contains other terms, particularly the anisotropic
dipole-dipole interaction, that breaks explicitly this $O(3)$ symmetry of the NLSM. This effective
energy functional thus produced
upon minimization a different set of solutions depending on the geometry of the trap in which the condensate is confined.

Given these
properties we shall now first see what type of approximations gives us this extended NLSM description of the dipolar condensate from the full second quantized Hamiltonian.
The full second quantized  Hamiltonian that incorporates effect of the dipole-dipole
interaction is given as:
%\begin{widetext}
\begin{eqnarray}
%\lefteqn{H} \nonumber \\
\lefteqn{H
= \int d{\bm r}_i \Biggl \{
{\hat \psi}_{m_{_1}}^\dagger ({\bm r}_i)
\left ( - \frac{\hbar^2}{2m} \nabla^2 + V_{\rm trap} ({\bm r}_i) - \mu \right )
{\hat \psi}_{m_{_1}} ({\bm r}_i)} \nonumber \\
&& \mbox{} +\sum_{F = 0}^{2f}
{\mathcal G}_{F} \langle m_{_1} m_{_2} | {\mathcal P}_F | m_{_3} m_{_4} \rangle
{\hat \psi}_{m_{_1}}^\dagger ({\bm r}_i)
{\hat \psi}_{m_{_2}}^\dagger ({\bm r}_i)
{\hat \psi}_{m_{_3}} ({\bm r}_i)
{\hat \psi}_{m_{_4}} ({\bm r}_i) \Biggr \}\nonumber \\
&& \mbox{} + \frac{c_{dd}}{2}
\int d{\bm r}_i \int d{\bm r}_j
\Biggl \{
{\hat \psi}_{m_{_1}}^\dagger ({\bm r}_i)
{\hat \psi}_{m_{_2}}^\dagger ({\bm r}_j)
{\hat \psi}_{m_{_3}} ({\bm r}_j)
{\hat \psi}_{m_{_4}} ({\bm r}_i) \nonumber \\
&& \mbox{} \times \frac{{\bm F}_{m_{_1} m_{_4}}
\cdot {\bm F}_{m_{_2} m_{_3}}
- 3 ({\bm F}_{m_{_1} m_{_4}}
\cdot {\vec e}_{ij})
({\bm F}_{m_{_2} m_{_3}}
\cdot {\vec e}_{ij})}
{r_{ij}^3} \Biggr \},
\label{eq:exacthamiltonian}
\end{eqnarray}
%\end{widetext}
In the above Hamiltonian we  have used Einstein summation convention implying that repeated indices stand for  summation.
The first expression in the above equation represents the usual single body
term where $\mu$ is chemical potential and $V_{\text{trap}}$ is the external
trapping potential. The second expression stands for the short-range
interaction between two atoms where $F$ is the hyperfine spin of each atom.
The coupling constant $\mathcal{G}_F$, which  corresponds to the scattering
to a state with  the total spin $F$,
is given by $\mathcal{G}_F = 4 \pi \hbar^2 a_S / m$ with an $s$-wave
scattering length $a_F$ and the mass of an atom $m$. Also, we set
${\vec e}_{ij} = ({\bm r}_i - {\bm r}_j) / r_{ij}$ and $r_{ij} =
|{\bm r}_i - {\bm r}_j|$.
$\mathcal{P}_F$ is the projection operator which projects the pair of
``1" and ``2"
atoms into a state with total hyperfine spin $F$. The operator can be
written using the Clebsch-Gordan
coefficient $\langle m_{_1} m_{_2} | F M_f \rangle$;
%\begin{eqnarray}
%\lefteqn{\langle m_{_1} m_{_2} | \mathcal{P}_S | m_{_3} m_{_4} \rangle}
%\nonumber \\
%&=& \sum_{M_s = -S}^S \langle m_{_1} m_{_2} | S Ms \rangle
%\langle S Ms | m_{_3} m_{_4} \rangle,
%\end{eqnarray}
\beq \langle m_{_1} m_{_2} | \mathcal{P}_F | m_{_3} m_{_4} \rangle = \sum_{M_f = -F}^F \langle m_{_1} m_{_2} | F M_f \rangle \langle F M_f | m_{_3} m_{_4} \rangle, \eeq
where $m_i$ is the spin sublevel of each atom, $F$ is the total spin of two
atoms,
$M_f$ is the spin sublevel of two atoms.
%And,
%sum over all possible spin sublevels.
The third term represents the long-ranged, anisotropic d-d interaction  with ${\bm F}$ is given by
%\begin{eqnarray}
%g_{_F} \mu_{_B} {\bm F}_{m_i m_j}
%&=& \langle m_i | {\hat {\bm \mu}} | m_j \rangle.
%\end{eqnarray}
$g_{_F} \mu_{_B} {\bm F}_{m_i m_j}
= \langle m_i | {\hat {\bm \mu}} | m_j \rangle$.
Here ${\hat {\bm \mu}} = - g_{_S} \mu_{_B} {\hat {\bm S}} + g_{_I} \mu_{_N}
{\hat {\bm I}}$, $\mu_{_N}$ is nuclear magneton, ${\hat {\bm S}}$ (${\hat {\bm I}}$)
is electronic (nuclear) spin angular momentum, and $g_{{_S}, {_I}}$ is Land{\' e} $g$
factors. Also, we set $c_{dd} = \mu_{_0} g_{_F}^2 \mu_{_B}^2 / (4 \pi)$ with
$\mu_{_0}$ is magnetic permeability of the vacuum.
Since the d-d interaction strength is proportional to $F^2$, it becomes more important for a large $F$ system.
%where the spin-spin interaction is much more complicated and bring us to very large parameter space.
Moreover in such a system, $F + 1$ contact interaction channels also exist. For these reasons, the analysis of such large $F$ system is very difficult and quite often hides the beauty of the quantum phase structures of such model under mathematical complexities.

Our purpose is to understand the effect of d-d interaction on rotationally invariant spin-textures. This requires us to understand the competition between rotationally invariant $s$-wave
contact interaction and the anisotropic d-d interaction. To this purpose we have considered the case where all the interaction channels that represent the short range interaction except the one that accounts the $s$-wave repulsive interaction  can be neglected in comparison to the d-d interaction. To motivate such a limit let us consider the case of
the $F = 1$  spinorial BEC \cite{ohmi,ho}. Here the interaction is characterized by
two scattering lengths  $a_0$ and $a_2$,  leading to the
 spin independent repulsive interaction $g \equiv g_0 = (\mathcal{G}_0 + 2 \mathcal{G}_2) / 3 = 4 \pi \hbar^2 (a_0 + 2 a_2) / 3m$ in the triplet channel and the spin dependent exchange interaction $g_2 = (\mathcal{G}_2 - \mathcal{G}_0) /3 = 4 \pi \hbar^2 (a_2 - a_0) /3m$ in the singlet channel.
Since $a_0$ and $a_2$ are comparable in the typical experiments, $g_2$ is actually much smaller than $g_0$;
$|g_2|/g\sim 1/10$ for $^{23}$Na \cite{stenger,burke} and $\sim 1/35$
for $^{87}$Rb \cite{barrett,klausen}.

This tendency that, except for the dominant repulsive part $g$,
other spin-dependent channels are nearly cancelled,
is likely to be correct for other $F$'s \cite{hirano} as well.
Therefore for larger $F$, it is expected that the d-d interaction should  become more important than these spin-spin interactions.  For example, in Chromium ($F=3$), it has been shown in reference \cite{Santos}
( see the equation $3$ of that paper) that the $s$-wave contact interaction in various spin channels is subdominant
as compared to the dominant repulsive $s$-wave interaction. Where as the relative strength of the d-d interaction goes up due to higher value of the $F$ as well as $c_{dd}$ as well as $F$.
Thus, to understand the interplay between the
contact interaction and the dipole dipole interaction even for chromium BEC one can begin with
only the dominant repulsive part of this short range interaction.

Once we accept this approximation,
instead of working with $\vec{\Psi} ({\bf r})$ composed of the full quantum mechanical
$2F + 1$ components $( \Psi_{F}, \Psi_{F-1}, \cdots, \Psi_{-F})$, the order parameter
can be simplified to $\vec{\Psi} ({\bf r}_i) =
\psi({\bf r}_i) {\vec S}({\bf r}_i)$
where ${\vec S}({\bf r}_i)$ is a classical vector with $| {\vec S} ({\bf r}_i)|^2 = 1$.
Namely we can treat it as the classical spin vector whose
magnitude $|\psi({\bf r}_i)|^2$ is proportional to the local condensate density.
As we see below such an approximation drastically simplify the Hamiltonian functional
that represents such spinorial dipolar system while still yielding a rich variety of quantum phases that may observed in such systems as static configurations at various values of the system parameters.
Naturally, larger
the spin, better will be this approximation.
Thus we investigate the stationary  states of the system  with large dipole moment using a classical spin approximation. Namely, our approximation is justified when  $[{\hat s}_i, {\hat s}_j] = i \epsilon_{i, j, k} \hbar {\hat s}_k / S_0 \ll 1$, where ${\hat s}_i$ is spin operator and $S_0$ is the magnitude of the spin. However, the current theory may also be applied to the spinor BECs with smaller value of spin, where our theory reproduces certain results similar to other studies \cite{kawaguchi}. This is very similar to the spin wave theory which should be valid for large spin only, but, works for spin $\frac{1}{2}$ particles as well.

 At this point it is also appropriate to point out the limitations of such theory.
For example, spin textures that arises out due to the effect of dipole dipole interaction on an anti ferromagnetic spin textures cannot be studied within this model since there is no term in the effective energy term ( without the d-d interaction) that favors the anti ferromagnetic ground state. We have completely neglect the terms that arises out of considering commutators of various spin operators as well. This condition is also exclude the possibility of observing the states which arises out of strong quantum fluctuations in spinor order parameter about it's mean field value. We shall point out some of the structures that appeared in the work of \cite{kawaguchi}, \cite{pu}, \cite{Santos,diener}, will not occur in this current theoretical framework.

The model Hamiltonian we propose is given by
%\begin{eqnarray}
%H
%&=& \int d^3{\bm r}_i \vec{\Psi}^\dagger ({\bm r}_i)
%H_0 ({\bm r}_i)
%\vec{\Psi} ({\bm r}_i) \nonumber \\
%%&& + \frac{1}{2} g_d \int d{\bm r}_i \int d{\bm r}_j
%&& \mbox{} + \frac{g_d}{2} \int \int d^3{\bm r}_i d^3{\bm r}_j
%V_{dd} ({\bm r}_i, {\bm r}_j)
%|\psi ({\bm r}_i)|^2 |\psi ({\bm r}_j)|^2, \nonumber \\
%\end{eqnarray}
\begin{eqnarray}
\lefteqn{H = \int d^3{\bm r}_i \vec{\Psi}^\dagger ({\bm r}_i)
H_0 ({\bm r}_i)
\vec{\Psi} ({\bm r}_i)} \nonumber \\
&& \mbox{} + \frac{g_d}{2} \int \int d^3{\bm r}_i d^3{\bm r}_j
V_{dd} ({\bm r}_i, {\bm r}_j)
|\psi ({\bm r}_i)|^2 |\psi ({\bm r}_j)|^2,
\end{eqnarray}
where
\vspace{-5mm}
\begin{eqnarray}
H_0
&=& - \frac{\hbar^2}{2 m} \nabla_i^2
+ V_{\rm trap} ({\bm r}_i)
- \mu + \frac{g}{2} | \vec{\Psi} ({\bm r}_i) |^2,
\end{eqnarray}
\begin{eqnarray}
V_{dd} ({\bm r}_i, {\bm r}_j)
&=& \frac{\vec{S}_i \cdot \vec{S}_j
- 3 (\vec{S}_i \cdot \vec{e}_{ij})
(\vec{S}_j \cdot \vec{e}_{ij})}
%{|{\bm r}_i - {\bm r}_j|^3}
{r_{ij}^3}.
\end{eqnarray}
The order parameter is represented by  the spinor ${\vec \Psi} ({\bm r}_i) = \psi ({\bm r}_i) {\vec S}_i$.
The uniaxially symmetric trap potential is given by $V_{\rm trap} ({\bm r}) = \frac{1}{2} m \omega^2 \{\gamma (x^2 + y^2)+z^2\}$, where $\omega$ is the trap frequency in the transverse direction and $\gamma$ the anisotropy parameter.
The short-range and d-d interaction parameters are introduced as $g$ and $g_d$, respectively. As explained earlier, the interactions that change each spin sublevel, are ignored
since they are subdominant to the d-d interactions and only spin-independent short-range interactions are considered with coupling constant $g$.
To implement the constraint  $|{\vec S}|^2=1$ we parametrize it as :
%\begin{eqnarray}
%{\vec S}_i
%\equiv {\vec S} ({\bm r}_i)
%= \left ( \begin{array}{c}
%\cos \varphi ({\bm r}_i) \sin \theta ({\bm r}_i) \\
%\sin \varphi ({\bm r}_i) \sin \theta ({\bm r}_i) \\
%\cos \theta ({\bm r}_i)
%\end{array} \right ).
%\end{eqnarray}
\beq {\vec S}_i \equiv {\vec S} ({\bm r}_i) = \begin{bmatrix} \cos \varphi ({\bm r}_i) \sin \theta ({\bm r}_i) \\
\sin \varphi ({\bm r}_i) \sin \theta ({\bm r}_i) \\
\cos \theta ({\bm r}_i)  \end{bmatrix}. \eeq
The magnitude of the spin is included in the d-d interaction coupling constant $g_d$.
The dimensionless form of this Hamiltonian functional of  $\psi ({\bm r}_i)$, $\theta ({\bm r}_i)$, $\varphi ({\bm r}_i)$ is
%\begin{widetext}
\begin{eqnarray}
\lefteqn{H
= {1\over 2}\int d^3{\bm r}_i \Biggl [ | \nabla \psi ({\bm r}_i) |^2
+ n_i
\Bigl \{ \bigl (\nabla \theta ({\bm r}_i) \bigr )^2} \nonumber \\
&& \mbox{} + \sin^2 \theta ({\bm r}_i)
\bigl ( \nabla \varphi ({\bm r}_i) \bigr)^2
+\gamma^2 (x^2 + y^2) + z^2\Bigr \} - 2 \mu n_i
+ g n_i^2
\Biggr ] \nonumber \\
&& \mbox{} + \frac{g_d}{2} \int \int d^3{\bm r}_i d^3{\bm r}_j
\left ( \frac{f ({\bm r}_i, {\bm r}_j)}{r_{ij}^3}
- 3 \frac{F ({\bm r}_i, {\bm r}_j)}{r_{ij}^5} \right )
n_i n_j,
\label{eq:exhamiltonian}
\end{eqnarray}
%\end{widetext}
where
\begin{eqnarray}
\lefteqn{f ({\bm r}_i, {\bm r}_j)
= \cos (\varphi_i - \varphi_j) \sin \theta_i \sin \theta_j
+ \cos \theta_i \theta_j,} \\
%\end{eqnarray}
%\vspace{-5mm}
%\begin{eqnarray}
\lefteqn{F ({\bm r}_i, {\bm r}_j)
= x_{ij}^2 \cos \varphi_i \cos \varphi_j \sin \theta_i \sin \theta_j}
\nonumber \\
&& \mbox{} + y_{ij}^2 \sin \varphi_i \sin \varphi_j \sin \theta_i \sin \theta_j
%\nonumber \\
%&& \mbox{}
+ z_{ij}^2 \cos \theta_i \cos \theta_j
\nonumber \\
&& \mbox{} + 2x_{ij} y_{ij} \sin (\varphi_i + \varphi_j) \sin \theta_i \sin \theta_j
%\nonumber \\
%&& \mbox{}
+ 2 y_{ij} z_{ij} \sin \varphi_i \sin \theta_i \cos \theta_j
\nonumber \\
&& \mbox{} + 2 z_{ij} x_{ij} \cos \varphi_i \sin \theta_i \cos \theta_j,
\end{eqnarray}
with $|\psi ({\bm r}_i)|^2 = |\psi_i|^2 = n_i$, $\theta_i = \theta ({\bm r}_i)$, $\varphi_i = \varphi ({\bm r}_i)$, and $\alpha_{ij} = \alpha ({\bm r}_i) - \alpha ({\bm r}_j)$ $(\alpha = x, y, z)$.
We note that the spin gradient term in the first line in Eq.~(\ref{eq:exhamiltonian}) represents the nonlinear sigma model \cite{nlsm,haldane} with the corresponding spin stiffness is given by
superfluid density. Thus the gradient energy term  gives the leading order  coupling between the spin with mass density and leads to
a number of interesting phenomenon in such condensates. Higher order coupling between spin and density
comes from the d-d interaction.
The  d-d interaction represents interaction between the different parts of the spin density.
Even though the rotational invariance of the NLSM is lost due to the explicit appearance of angular fields $\phi$ and $\theta$ in the energy density, the system still have
reflection symmetry in the $xy$,$yz$ and $zx$ plane. In the subsequent analysis we shall see that this reflection symmetry is manifested in the equilibrium configurations of the condensates.
 The energy (length) is scaled by the harmonic frequency $\omega$ (harmonic length $d = 1/ \sqrt{m \omega}$ ) with $\hbar = 1$. The functional derivatives with respect to $\psi ({\bm r}_i)^\ast$, $\theta ({\bm r}_i)$, and $\varphi ({\bm r}_i)$ lead to the corresponding the time-dependent GP equations in the imaginary time $\tau$:
\begin{eqnarray}
\lefteqn{\frac{\partial \psi ({\bm r}_i, \tau)}{\partial \tau}%} \nonumber \\
= - \frac{\delta H}{\delta \psi^\ast ({\bm r}_i, \tau)}} \nonumber \\
&=& \frac{1}{2} \nabla_i^2 \psi_i + \left ( \mu - g n_i \right ) \psi_i
\nonumber \\
%&& \mbox{} - \frac{1}{2} \bigl \{
%(\nabla_i \theta_i)^2 + \sin^2 \theta_i (\nabla_i \varphi_i)^2 \nonumber \\
%&& \mbox{} + \gamma^2 (x_i^2 + y_i^2) + z_i^2 \bigr \} \psi_i
%\nonumber \\
&& \mbox{} - \frac{1}{2} \left \{
(\nabla_i \theta_i)^2 + \sin^2 \theta_i (\nabla_i \varphi_i)^2
+ \gamma^2 (x_i^2 + y_i^2) + z_i^2 \right \} \psi_i \nonumber \\
&& \mbox{} - g_d \int d{\bm r}_j
\left ( \frac{f ({\bm r}_i, {\bm r}_j)}{r_{ij}^3}
- 3 \frac{F ({\bm r}_i, {\bm r}_j)}{r_{ij}^5} \right ) n_j \psi_i,
\label{eq:psigp}
\end{eqnarray}
\begin{eqnarray}
\lefteqn{\frac{\partial \theta ({\bm r}_i, \tau)}{\partial \tau}%} \nonumber \\
= - \frac{\delta H}{\delta \theta ({\bm r}_i, \tau)}} \nonumber \\
&=& (\nabla_i^2 \theta_i) n_i + (\nabla_i \theta_i) \cdot (\nabla_i n_i)
- \sin \theta_i \cos \theta_i (\nabla_i \varphi_i)^2 n_i \nonumber \\
&& \mbox{} - g_d \int d{\bm r}_j \{
\cos (\varphi_j - \varphi_i) \sin \theta_j \cos \theta_i% \nonumber \\
%&&\mbox{}
- \sin \theta_i \cos \theta_j \} \frac{n_i n_j}{r_{ij}^3}
\nonumber \\
&& \mbox{} + 3 g_d \int d{\bm r}_j \{
x_{ij}^2 \sin \theta_j \cos \theta_i \cos \varphi_i \cos \varphi_j \nonumber \\
&& \mbox{}
+ y_{ij}^2 \sin \theta_j \cos \theta_i \sin \varphi_i \sin \varphi_j
- z_{ij}^2 \sin \theta_i \cos \theta_j \nonumber \\
&& \mbox{}
+ x_{ij} y_{ij} \sin \theta_j \cos \theta_i \sin (\varphi_i + \varphi_j)
\nonumber \\
&& \mbox{}
+ y_{ij} z_{ij} (\cos \theta_j \cos \theta_i \sin \varphi_i
- \sin \theta_j \sin \theta_i \sin \varphi_j) \nonumber \\
&& \mbox{}
+ z_{ij} x_{ij} (\cos \theta_j \cos \theta_i \cos \varphi_i
- \sin \theta_j \sin \theta_i \cos \varphi_j)% \nonumber \\
%&& \mbox{}
\} \frac{n_i n_j}{r_{ij}^5},
\label{eq:thetagp}
\end{eqnarray}
%\vspace{-5mm}
and
\begin{eqnarray}
\lefteqn{\frac{\partial \varphi ({\bm r}_i, \tau)}{\partial \tau}%} \nonumber \\
= - \frac{\delta H}{\delta \varphi ({\bm r}_i, \tau)}} \nonumber \\
&=& \sin^2 \theta_i (\nabla_i^2 \varphi_i)
+ 2 \sin \theta_i \cos \theta_i (\nabla_i \theta_i)
\cdot (\nabla_i \varphi_i) n_i \nonumber \\
&& \mbox{} + \sin^2 \theta_i (\nabla_i \varphi_i) \cdot (\nabla_i n_i)
\nonumber \\
&& \mbox{} - g_d \int d{\bm r}_j
\sin (\varphi_j - \varphi_i) \sin \theta_j \sin \theta_i
\frac{n_i n_j}{r_{ij}^3} \nonumber \\
&& \mbox{} + 3 g_d \int d{\bm r}_j \{
- x_{ij}^2 \sin \theta_j \sin \theta_i \sin \varphi_i \cos \varphi_j
\nonumber \\
&& \mbox{}
+ y_{ij}^2 \sin \theta_j \sin \theta_i \sin \varphi_j \cos \varphi_i
\nonumber \\
&& \mbox{}
+ x_{ij} y_{ij} \sin \theta_j \sin \theta_i \cos (\varphi_i + \varphi_j)
\nonumber \\
&& \mbox{}
+ y_{ij} z_{ij} \sin \theta_i \cos \theta_j \cos \varphi_i \nonumber \\
&& \mbox{}
- z_{ij} x_{ij} \cos \theta_j \sin \theta_i \cos \varphi_i
\} \frac{n_i n_j}{r_{ij}^5}.
\label{eq:phigp}
\end{eqnarray}
These equations describe the time evolution of order parameters for a ``imaginary time" $\tau$, which rolls along the slope of the energy functionals. For $\tau \rightarrow \infty$, the order parameters converge to the stationary configuration, corresponding to one of the local minima of the energy functional.
We numerically solved the set of the GP equations (\ref{eq:psigp})-(\ref{eq:phigp}) in imaginary time ($\tau = i t$).% and found several stationary states. Those are understandable by considering anisotropy of the d-d interaction and spin gradient term.
In this paper, for a  fixed repulsive interaction $(g/\omega d^3 = 0.01)$, we vary the d-d interaction $g_d$ in a range of $0 \le g_d \le 0.4 g$, beyond which the system is unstable. We consider two types of the confinement to see the effect of of the trap anisotropy on the d-d interaction: namely a cigar $(\gamma = 5.0)$ and a pancake $(\gamma = 0.2)$ trap . The total particle number is kept around $N \sim 10^4$. The three-dimensional space is discretized into the lattice sites $\sim 2.5 \times 10^4$.
%\section{DIPOLE-DIPOLE INTERACTION}
\section{STABLE CONFIGURATIONS \label{sec:result}}
\begin{figure}[ht]
%\centerline{ \epsfxsize 5cm \epsfysize 4cm \epsffile{F1.eps} }
\centerline{ \epsfxsize 0.5\linewidth \epsffile{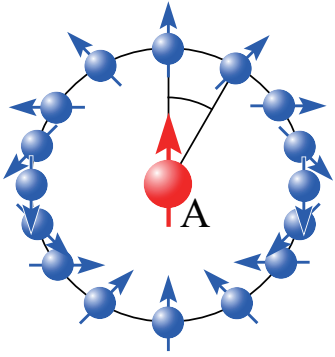} }
\caption{{\it color online}: Anisotropy of the d-d interaction considering two spins. The spins around the spin A at the center are pointing the direction of the d-d energy minimum. The top and bottom spins are likely to be parnell to the spin A or head-to-tail texture, and the spins at the side of spin A are likely to be antiparallel to the spin A.}
\label{fig:d-dint}
\end{figure}
\begin{figure}[ht]
%\centerline{\epsffile{spin.eps}}
%\centerline{ \epsfxsize 8cm \epsfysize 7cm \epsffile{spin.eps}}
\centerline{ \epsfxsize \linewidth \epsffile{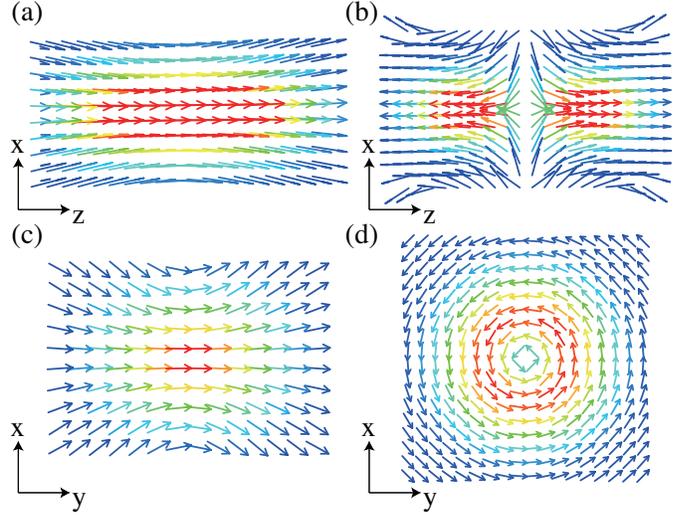}}
\caption{{\it color online}: Spin texture $\bm{S}(\bm{r})$ of the stationary states. Four figures are correspond to (a) z-flare texture, (b) two-z-flare texture, (c) r-flare texture, and (d) spin current texture. (a) and (b) are in the cigar type trap potential $\gamma \! = \! 5$. (c) and (d) are in the pancake type trap potential $\gamma \! = \! 0.2$. The d-d interaction strength are all the same value $g_d \! = \! 0.2 g$. The color is corresponds to the local density.}
\label{fig:spin}
\end{figure}
\begin{figure}[ht]
\centerline{ \epsfxsize \linewidth \epsffile{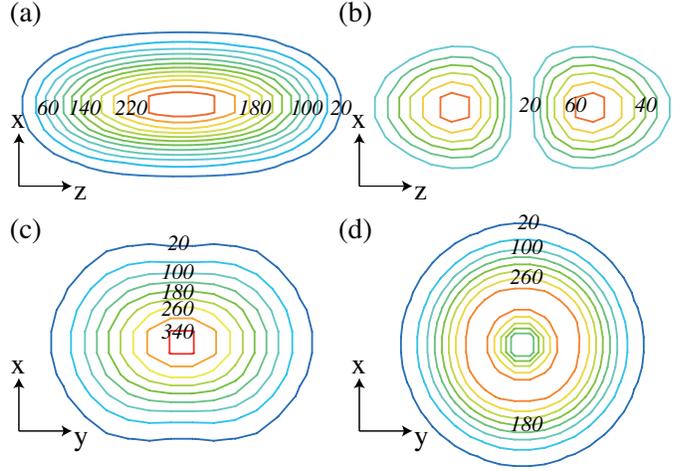} }
\caption{{\it color online} Density profile of the stationary states. Four figures are correspond to (a) z-flare texture, (b) two-z-flare texture, (c) r-flare texture, and (d) spin current texture. (a) and (b) are in the cigar type trap potential $\gamma \! = \! 5$. (c) and (d) are in the pancake type trap potential $\gamma \! = \! 0.2$. The d-d interaction strength are all the same value $g_d \! = \! 0.2 g$. In the spin current texture (d), the density is torus structure. The maximum point of the density are not at the center in (b) and (d), due to their spin structures.}
\label{fig:density}
\end{figure}

First, let us consider the anisotropic nature  of the d-d interaction.
Figure \ref{fig:d-dint} shows the anisotropy of the d-d interaction with the help of  two spins (one at the center and the another is taken around it). The directions of the surrounding correspond to the  energy minimum. Those spins at the top and the bottom are likely to be parallel to the spin A or head-to-tail texture, and the spins at both sides of spin A are anti-parallel to the spin A.
Now we shall describe a number of equilibrium  spin-density configuration that are obtained
by minimizing the energy expression (\ref{eq:exhamiltonian}). In Fig. \ref{fig:spin} (a) and Fig. \ref{fig:density} (a) we
have respectively plotted the spin-profile of the z-flare  structure as well as its density profile in the $xz$ plane which is obtained in a cigar shaped trap
($\gamma=5$). This spin texture is the typical example of the dipolar effect on an otherwise ferromagnetic structure, that is the ground state of isotropic NLSM. At the trap center, where the density is the highest, we can have a spin vector (dipole moment) pointing toward $z$ direction. And the other spins placed right and left are laid in the head-to-tail texture. The spins near the surface of the system are bent as those in Fig.~\ref{fig:d-dint}. As $g_d$ increases ( $= 0.1 g$ and $0.2 g$), the bending angle of the flare spin texture increases and the system gets elongated along the $z$ direction.
The typical length scale over the change of the spin angle takes place is determined by the relative strength of the gradient energy  term to the term representing the d-d interaction. At
the center, the spin-stiffness or the superfluid density is higher and hence bending the spin is energetically costly. Thus the energy will be minimized by reducing the gradient energy. At the edge
the spin-stiffness is lower and and hence the bending of the spin will cost less gradient energy.
The texture in Fig.~\ref{fig:spin}(c) is another example that shows strong coupling of the density and the spin texture and it also  breaks the axial symmetry of the density in the $x$-$y$ plane. The texture is the so-called ``r-flare" texture, and is one of the minimum energy configurations in a disk-shaped pancake geometry, namely when $\gamma \ll 1$. Because of the stronger confinement along the $z$-axis the  gradient
energy cost will be higher in that direction. Therefore the flaring will now take place in $x$-$y$
plane, rather than in the $x$-$z$ plane. Let us assume,
the spins are aligned to one direction in the $x$-$y$ plane (e.g., the $x$ direction in Fig.~\ref{fig:spin}(c).  Fig.~\ref{fig:density}(c) shows the cross-section density plots of the particle number (the corresponding spin texture). The spins are almost parallel to the $x$ axis, but they bend away at the outer region. At the $x$-$y$ plane around $z = 0$, the condensate gets elongated to the direction along the spin polarization. This is
because of the d-d interaction, which favors the head-to-tail arrangement, and the particle number is increased at the center. The bending angle of the spins on the circumference increases with increasing $g_d$ in the same way it happens in the case of $z$-flare configuration.
The two structure that has been discussed above are non-singular in terms of density as well as
the spin-profile even though they have shown strong intertwining between the spin and density. We shall now discuss two cases where this intertwining lead to singularity in either density or spin-profile or in the both.
The spin and density profile  of one of such equilibrium structure  is shown in Fig.~\ref{fig:spin}(b) and Fig.\ref{fig:density} (b).
This configuration is obtained from the initial condition of a hedgehog texture like topological defect in which all the spins point outward from the origin, i.e., ${\bf S} ({\bm r}) = {\bm r} / |{\bm r}|$. The
solution presented here can also be regarded as the two ``z-flare" textures stacked back to back. Here, the density at the trap center is decreased to avoid spin gradient energy cost even though this region represents the minimum of the trap potential and finally leads to the splitting of the
condensate. This clearly demonstrates the strong-coupling between spin and density. The leading order contribution comes from the spin gradient energy whose coefficient or stiffness is
proportional to the density of the condensate. But the dipole-dipole interaction term also contributes to this process since it includes all the higher order derivatives of the spinorial fields.
The final configuration which we shall discuss in this context is the so-called ``spin current" texture. This structure combines a vortex in the
density field with a  circulation current in the spin-texture.
Thus this serves as an another typical configuration in which the spin texture and the particle density are strongly coupled through the d-d interaction. Symmetrically on the both side
of the density depleted region, the spins are locally aligned to head-to-tail texture and the whereas spins are antiparallel (see  Fig. \ref{fig:spin}(d)). This texture is explained by the anisotropy of the d-d interaction depicted in Fig.~\ref{fig:d-dint}. At the trap center, the density depletion is occurred over the coherent length $\xi_d$ of the d-d interaction which is
shown in Fig. \ref{fig:density}(d). In the present case, we find $\xi_d \sim 2.0 \xi_c$ ($\xi_c$ is the ordinary coherent length of the repulsive interaction). This is because to avoid spin gradient energy cost.
For an alternative explanation of the spin current texture we rewrite the d-d interaction in Eq.~(\ref{eq:exhamiltonian}) as $V_{dd} ({\bm r}_{ij}) \propto (g_d/r_{ij}^3) \sum_{\mu = -1}^2 Y_{2 \mu} (\cos \theta) \sum_\mu (i j)$ with $\sum_\mu (i j)$ being a rank 2 tensor consisting of the two spins at $i$ and $j$ sites, and $Y_{2 \mu} (\cos \theta)$ a spherical harmonics \cite{pethick}. $\theta$ is the polar angle in spherical coordinates of the system. The spin current texture shown in  Fig.~\ref{fig:spin}(d) picks up the phase factor $e^{2 i \phi}$ when winding around the origin. This is coupled to $Y_{2 \pm 2} (\cos \theta) \propto \sin^2
\theta$, meaning that this orbital moment dictates the number density depletion at the pancake center. The spin-orbit coupling directly manifests itself here. The total angular momentum consisting of the spin and orbit ones is a conserved quantity of the present axis-symmetric system, leading to the Einstein-de Haas effect \cite{kawaguchi}. The spin current texture is stable for the wide range of anisotropy $\gamma: 0.01 \le \gamma \le 0.6$, beyond which it becomes unstable.
Some of the spin textures are similar to those obtained by Kawaguchi {\it et al.} \cite{kawaguchi} for $F = 1$ case, such as the flare (the flower in their terminology) and the spin current textures (the polar core vortex). The latter texture demonstrates the interaction between the orbital and spin angular momenta, ultimately leading to the Einstein-de Haas effect mentioned in connection with $F = 3$ $^{52}$Cr case \cite{stuhler} and $F = 1$ case \cite{kawaguchi}.
\section{STABILITY REGIME FOR VARIOUS QUANTUM  PHASES \label{sec:phases} }
In the previous section \ref{sec:result} we have described how energy minimization process in presence
of dipole-dipole interaction takes advantage of the strong coupling between mass and spin-density and
stabilizes a number of topologically non trivial structure. All these structures correspond to various quantum mechanical ground states of the corresponding many body systems. The transition of the system from one macroscopic configuration to another as a function of various external parameters such as the coupling constant $\frac{g_d}{g}$ that gives the relative strength of dipolar interaction, anisotropy of the trap $\gamma$, and the chemical potential $\frac{E}{N}$, actually implies quantum phase transition when applied to infinite system. Indeed the system on which we perform our numerical computation is finite and results are also limited by the numerical accuracy. Moreover we only have been
able to focus on some limited data points in the three dimensional phase space of external parameters as described above. Nevertheless based on this we can provide some indication of interesting quantum phase transitions or cross-over in the actual system. We make a few comments in this direction and then summarize our numerical
results in a tabular form.
For example the r-flare and the spin-current texture both are stabilized in a pancake geometry. To
see what external condition energetically prefers one above the other, we have looked at the how
the energy $E$ varies as a function of the total number of particles for a given aspect ratio is $\gamma = 0.2$ and coupling strength $g_d = 0.2 g$. We found that
in the small $N$ region, the r-flare texture is favorable, but in the larger $N$ region, the spin current texture has the lower energy. The z-flare texture is  never the ground state even though sometimes it corresponds to a local energy minima at these values of the parameters. These can be explained as follows: At the small $N$ region, the system size is comparable with the coherence length $\xi_d$ of the d-d interaction. Thus it costs more spin gradient energy than those in large $N$. The z-flare texture can not be the lowest energy state  because the alignment and elongation to $z$ direction cost the higher energy than those to the radial direction at this value of the trap anisotropy (aspect ratio).
The $z$-flare spin texture is however stable for $g_d < 0.3 g$ and for $N \sim 2.3 \times 10^4$, and stay robust for various aspect ratios such as: $\gamma = 0.2$ and $5.0$, but it corresponds to the minimum energy
configuration only in a cigar shaped trap $\gamma \gg 1$.
Again, in this regime $r$-flare structure is stable, but has higher energy compared to the z-flare structure. This also points out the another aspect of this
way of computing the ground state energy. The stable solutions may correspond to one of the local minima of the energy landscape and  thus a comparison between the energies corresponding to various such local minima
is necessary to understand the actual equilibrium structure. As we increase the coupling strength beyond
$g_d =0.4g$ for the particle  number of order $10^4$, none of the above mentioned structure is stable and due
to attractive part of the d-d interaction the condensate collapse. However this value of $g_d$ is typically higher than the critical interaction strength  where all dipoles are polarized in a given direction.
In the following paragraph we put some of these results in a tabular form. Here \textbf{rf} means r-flare, \textbf{zf} means z flare, and \textbf{sc} means spin-current texture. The last column not necessarily indicate the ground state, but only an equilibrium structure within computational accuracy.
For example comparing the data in $8$th, $9$th, and
$10$th row of the table which are for the almost same number of particles we find that the \textbf{rf} structure is the energetically most favored structure. Similarly a comparison between the $8$th and $11$th row indicates that as one goes from a pancake to a cigar shaped trap the \textbf{zf} structure becomes energetically becomes more favorable.
\begin{table}
\caption{Energy of various stable configurations.}
\center{
\begin{tabular}{|c|c|c|c|}
\hline
$\gamma$ & $\frac{g_d}{g}$ & $\frac{E}{N}$  & \textbf{type} \\
\hline
$0.05$   & $0.20$  &  $1.535$   & \textbf{sc} \\
\hline
$0.20$   &  $0.10$  &  $1.855$  &  \textbf{zf} \\
\hline
$0.20$   &  $0.10$  &  $2.716$  &  \textbf{zf} \\
\hline
$0.20$   &  $0.10$  &  $1.617$  &  \textbf{rf} \\
\hline
$0.20$   &  $0.10$  &  $2.386$  &  \textbf{rf} \\
\hline
$0.20$   &  $0.10$  &  $1.569$  &  \textbf{sc} \\
\hline
$0.20$   &  $0.10$  &  $2.344$  &  \textbf{sc} \\
\hline
$0.20$ & $0.20$   & $2.308$   &  \textbf{zf} \\
\hline
$0.20$ & $0.20$   & $1.900$   &  \textbf{rf} \\
\hline
$0.20$ & $0.20$   & $1.988$   &  \textbf{sc} \\
\hline
$5.00$  & $0.29$   & $1.800$  &  \textbf {zf} \\
\hline
\end{tabular}
}
\end{table}
%As seeing those configuration, we notice that the spin texture and particle density are strongly coupled. This coupling make the density profile of the system anisotropic: elongate to the spin aligned direction and shortage the perpendicular to it. This is more clearly shown in Fig.~\ref{fig:rflare}.
%-----?t?I`?UN?UN?g??
%These spin textures can be observed directly via a novel phase-sensitive {\it in situ} detection \cite{review} or indirectly via conventional absorption imaging for the number density.
%It is interesting to examine the vortex properties under rotation.
%For the spin current texture, the vortex entry into a system should be easy because in the central region the mass density is already depleted. We point out that the collective modes might be also intriguing because the mass density is tightly coupled with the spin degrees of freedom. These problems belong to future work.
%----/
\section{ Theory of collective excitations \label{sec:collective}}
The next interesting issue will be to study the nature of collective excitations around such ground state configurations. The methodology is similar to the study of collective excitations in ordinary BEC, however now for these spinorial order parameter which represents strong intertwining between the density and the spin degrees of freedom. We can therefore expect that some of these resulting collective excitations strong coupling between density waves such as phonons with various types of spin waves. The method is outlined here briefly. Our order parameter consists of three field, namely
\beq \bs{\chi} = \begin{bmatrix} \chi_1 \\ \chi_2 \\ \chi_3 \end{bmatrix} = \begin{bmatrix} \psi \\ \theta \\ \varphi \end{bmatrix} \eeq
As GP solution, $\psi$ can be complex whereas the angular fields $\theta$ and $\varphi$ are real.
We consider the small fluctuations around each of this field of the form
\beq \chi_i \Rightarrow \chi_i + \lambda (u_i e^{ i \omega t} - v_i e^{-i \omega t}) \eeq
For $\{ v_{i} \}$ we take the complex conjugate.
We then subsequently inserts them in the field equations and in the spirit of linear response theory write the resulting equations by collecting the terms that appears as coefficients of
$\lambda$. We neglect all other terms which appears with higher power of $\lambda$. These equations are very long because of the presence of the $d-d$ interaction in the mean field energy functional. We therefore include the $d-d$ interaction term only symbolically and illustrate them in one case explicitly

The equations for the collective excitations are respectively:\\
For $\psi$

\bea \omega u_{1} & = & \frac{1}{2}[ \nabla^2 + 2 \mu -4g |\psi|^2 - (\nabla \theta)^2 -\sin^2 \theta (\nabla \varphi)^2 - V_{\rm trap}]u_1 \nonumber \\
& & \mbox{}-\frac{1}{2}\psi[(\nabla \theta) \nabla +\sin 2\theta  (\nabla \varphi)^2]u_2 -\frac{1}{2}\psi \sin^2 \theta (\nabla \varphi)\nabla u_3 \nonumber \\
& & \mbox{}+ g \psi^2 v_1 + \frac{1}{2}\psi (\nabla \theta)\nabla v_2 + \frac{1}{2} \psi \sin^2 \theta (\nabla \varphi)\nabla v_3 \nonumber \\
& & \mbox{} + \text{d-d}_{contrib} \label{u1eq} \eea

\bea \omega v_{1} & = & -\frac{1}{2}[ \nabla^2 + 2 \mu -4g |\psi|^2 - (\nabla \theta)^2 -\sin^2 \theta (\nabla \varphi)^2 - V_{\rm trap}]v_1 \nonumber \\
& & \mbox{}+\frac{1}{2}\psi^{*}[(\nabla \theta) \nabla +\sin 2\theta  (\nabla \varphi)^2]v_2 +\frac{1}{2}\psi^{*} \sin^2 \theta (\nabla \varphi)\nabla v_3 \nonumber \\
& & \mbox{} -g (\psi^{*})^2 u_1 - \frac{1}{2}\psi (\nabla \theta)\nabla u_2 + \frac{1}{2} \psi^{*} \sin^2 \theta (\nabla \varphi)\nabla u_3 \nonumber \\
&  & \mbox{} + \text{d-d}_{contrib}\label{v1eq} \eea

For $\theta$
\bea \omega u_{2} & = & [\psi^{*}(\nabla^2 \theta) +(\nabla \theta)((\nabla \psi^{*}) + \psi^{*} \nabla) -\frac{1}{2} \sin 2\theta (\nabla \varphi)^2 \psi^{*}]u_1 \nonumber \\
& & \mbox{} + [ |\psi|^2 \nabla^2 + (\nabla |\psi|^2)\nabla -\frac{1}{2}\cos 2 \theta (\nabla \varphi)^2 |\psi|^2]u_2 \
\nonumber \\
&  & \mbox{}-\frac{1}{2}\sin 2\theta (\nabla \phi) |\psi|^2 \nabla u_3 \nonumber \\
& & \mbox{}
-[(\nabla^2 \theta) \psi + (\nabla \theta) ((\nabla \psi) + \psi \nabla)
%\nonumber \\ & & \mbox{}
-\frac{1}{2}\sin 2\theta (\nabla \varphi)^2 \psi]v_1 \nonumber \\
& & \mbox{}+ \sin \theta \cos \varphi |\psi|^2 (\nabla \varphi) \nabla v_3  + \text{d-d}_{contrib} \label{u2eq} \eea

\bea \omega v_{2} & = & -[\psi(\nabla^2 \theta) +(\nabla \theta)((\nabla \psi) + \psi \nabla)) - \frac{1}{2}\sin 2\theta (\nabla \varphi)^2 \psi]v_1 \nonumber \\
& & \mbox{} - [ |\psi|^2 \nabla^2 + (\nabla |\psi|^2)\nabla -\frac{1}{2}\cos 2 \theta(\nabla \varphi)^2 |\psi|^2]v_2
\nonumber \\
&  & \mbox{}+\frac{1}{2}\sin 2\theta (\nabla \phi) |\psi|^2 \nabla v_3 \nonumber \\
& & \mbox{}+[(\nabla^2 \theta) \psi + (\nabla \theta) ((\nabla \psi^{*}) + \psi^{*} \nabla) -\frac{1}{2}\sin 2\theta  |\nabla \varphi|^2 \psi]u_1\nonumber \\
& & \mbox{}+ \sin \theta \cos \varphi |\psi|^2 (\nabla \varphi) \nabla u_3 + \text{d-d}_{contrib}  \label{v2eq} \eea
For $\phi$
\bea \omega u_3 & = &[\sin 2\theta (\nabla \phi)(\nabla \theta)\psi^{*}+\sin^2 \theta (\nabla \varphi)((\nabla \psi^{*}) + \psi^{*}\nabla)]u_1  \nonumber \\
& & \mbox{}+ [\frac{1}{2}\sin 2\theta [(\nabla^2 \varphi) +2(\nabla \varphi)[|\psi|^2 \nabla + (\nabla|\psi|^2)] \nonumber \\
& & \mbox{} +\cos 2 \theta(\nabla \theta)(\nabla \phi)|\psi|^2 ]u_2  \nonumber \\
& & \mbox{} + [\sin^2 \theta\{ |\psi|^2 \nabla^2 + (\nabla |\psi|^2)\nabla\} + \sin 2 \theta (\nabla \theta)|\psi|^2 \nabla]u_3 \nonumber \\
& & \mbox{} -[\sin 2 \theta(\nabla \theta)(\nabla \varphi)\psi + \sin^2 \theta (\nabla \varphi)((\nabla \psi) + \psi \nabla)]v_1 \nonumber \\
& & \mbox{}+ \text{d-d}_{contrib} \label{u3eq} \eea
\bea \omega v_3 & = &-[ \sin 2\theta(\nabla \phi)(\nabla \theta)\psi+\sin^2 \theta (\nabla \varphi)((\nabla \psi) + \psi\nabla)]v_1  \nonumber \\
& & \mbox{}- [\frac{1}{2}\sin 2\theta [(\nabla^2 \varphi) +2(\nabla \varphi)[|\psi|^2 \nabla + (\nabla|\psi|^2)] \nonumber \\
& & \mbox{} -\cos 2 \theta(\nabla \theta)(\nabla \phi)|\psi|^2 ]v_2  \nonumber \\
& & \mbox{} - [\sin^2 \theta\{ |\psi|^2 \nabla^2 + (\nabla |\psi|^2)\nabla\} + \sin 2 \theta (\nabla \theta)|\psi|^2 \nabla]v_3 \nonumber \\
& & \mbox{} + [\sin 2 \theta(\nabla \theta)(\nabla \varphi)\psi^{*} + \sin^2 \theta (\nabla \varphi)((\nabla \psi^{*}) + \psi^{*} \nabla)]u_1 \nonumber \\
& & \mbox{}+ \text{d-d}_{contrib} \label{v3eq} \eea
These gives us a set of six equations for six $u_i$s and $v_i$s. Their solutions describe the nature of the collective excitations. The $d-d$ interaction contribution to the first equations for this set (\ref{u1eq}) (after collecting the coefficients of $e^{i \omega t}$s only) will be for example
\bea \text{d-d}_{contrib}& = & -g_d \int d\bs{r_{i}} ( \frac{f_{ij}}{r_{ij}^3}-3\frac{F_{ij}}{r_{ij}^5})\psi_{j}(\psi_{i}^{*}u_{1i}+\psi_{i}v_{1i}) \nonumber \\
& & \mbox{}- g_d \int dr_{i}( \frac{f_{ij}}{r_{ij}^3}-3\frac{F_{ij}}{r_{ij}^5}) n_i u_{1j} \nonumber \\
& & \mbox{} -g_d\int dr_{i} \mathcal{F}_{ij} n_i \psi_{j} \eea
where
\bea \mathcal{F}_{ij} & = & \frac{1}{r_{ij}^3}[\{\cos(\phi_{i}-\phi_{j})\cos \theta_i \sin \theta_j -\sin \theta_i \cos \theta_j \}u_{2i} \nonumber \\
& & \mbox{} + \{\cos(\phi_{i}-\phi_{j})\sin \theta_i \cos \theta_j -\cos \theta_i \sin \theta_j \}u_{2j}
\nonumber \\
& & \mbox{}-\sin(\phi_i - \phi_j)\sin \theta_i \sin \theta_j (u_{3i}-u_{3j})] \nonumber \\
&  & \mbox{} -3\frac{1}{r_{ij}^5}[\{ x_{ij}^2\cos \theta_{i} \sin \theta_{j} \cos \phi_{i} \cos \phi_{j} \nonumber \\
& & \mbox{} +y_{ij}^2
\cos \theta_i \sin \theta_{j} \sin \phi_{i} \sin \phi_{j} \nonumber \\
& & \mbox{}- z_{ij}^2 \sin \theta_{i} \cos \theta_{j} \nonumber \\
& & \mbox{}+ x_{ij}y_{ij}\cos \theta_{i} \sin \theta_{j} \sin(\phi_{i}+\phi_{j}) \nonumber \\
& & \mbox{} + 2y_{ij}z_{ij}\cos \theta_{i} \cos \theta_{j} \sin \phi_{i} \nonumber \\
& & \mbox{}- 2z_{ij}x_{ij} \sin \theta_{i} \sin \theta_{j} \cos \phi_{i} \}u_{2i} \nonumber \\
&  & \mbox{} + \{ x_{ij}^2\sin \theta_{i} \cos \theta_{j} \cos \phi_{i} \cos \phi_{j} \nonumber \\
& & \mbox{}+ y_{ij}^2 \sin \theta_i \cos \theta_{j} \sin \phi_{i} \sin \phi_{j} \nonumber \\
& & \mbox{}- z_{ij}^2 \cos \theta_{i} \sin \theta_{j} \nonumber \\
& & \mbox{}+ x_{ij}y_{ij}\sin \theta_{i} \cos \theta_{j} \sin(\phi_{i}+\phi_{j}) \nonumber \\
& & \mbox{} - 2y_{ij}z_{ij}\sin \theta_{i} \sin \theta_{j} \sin \phi_{i} \nonumber \\
& & \mbox{} +2z_{ij}x_{ij} \cos \theta_{i} \cos \theta_{j} \cos \phi_{i} \}u_{2j} \nonumber \\
& & \mbox{}+ \{ -x_{ij}^2 \sin \theta_{i} \sin \theta_{j} \sin \phi_{i} \cos \phi_{j} \nonumber \\
& & \mbox{}+ y_{ij}^2 \sin \theta_{i} \sin \theta_{j} \cos \phi_{i} \cos \phi_{j} \nonumber \\
& & \mbox{} + x_{ij}y_{ij} \sin \theta_{i} \sin \theta_{j}\cos(\phi_{i}+\phi_{j}) \nonumber \\
& & \mbox{} + 2y_{ij}z_{ij}\sin \theta_{i} \cos \theta_{j} \cos \phi_{i} \nonumber \\
& & \mbox{}-2z{ij}x_{ij}\cos \theta_{i} \sin \theta{j} \sin \phi_{i} \}u_{3i} \nonumber \\
& & \mbox{} + \{-x_{ij}^2 \sin \theta_{i} \sin \theta_{j} \cos \phi_{i} \sin \phi_{j} \nonumber \\
& & \mbox{} + y_{ij}^2 \sin \theta_{i} \sin \theta_{j}\sin \phi_{i} \cos \phi_{j} \nonumber \\
& & \mbox{} + x_{ij}y_{ij} \sin \theta_{i} \sin \theta_{i} \sin \theta_{j} \cos(\phi_{i}+\phi_{j}) \}u_{3j}]
\eea

In the similar manner all other contributions can also be written
Full numerical solutions of these equations will be submitted for publication in a later communication. Here we therefore made some comments by pointing out the general nature of the collective excitations under the applications of a weak uniform field.
The simplest limit is the case without dipolar interaction
and the trapping potential. The fields are uniform. The ground state is an uniform ferromagnetic superfluid.
Thus $\nabla \psi = \nabla \theta = \nabla \varphi =0$

The above set of equations become

\bea \omega u_1 & = & [\frac{1}{2} \nabla^2 + \mu -2 g |\psi|^2]u_1 + g|\psi|^2 v_1 \nonumber \\
\omega v_{1} &  = & -[\frac{1}{2} \nabla^2 + \mu -2 g |\psi|^2]v_1 - g|\psi|^2 u_1 \label{psiex} \\
\omega u_2 & = & |\psi|^2 \nabla^2 u_2 \nonumber \\
\omega v_2 & = & -|\psi|^2 \nabla^2 v_2 \nonumber \label{thetaex}\\
\omega u_3 & = & |\psi|^2 \sin^2 \theta \nabla^2 u_3 \\
\omega v_3 & = & -|\psi|^2 \sin^2 \theta \nabla^2 v_3 \label{phiex} \eea

The first set of equations (\ref{psiex})  describes the usual Bogoliubov modes for the uniform condensate with
\beq \omega = \pm {\sqrt{\varepsilon_k(\varepsilon_k + 2g|\psi|^2)}} , \varepsilon_k = \frac{k^2}{2} \eeq
where we have used $\mu= g|\psi|^2$ for the uniform condensate.
In the long wavelength limit this gives
\beq \omega \propto k, k \ll 1 \eeq
the usual phonon mode. The two other equations (\ref{thetaex}) and (\ref{phiex}) gives
\beq \omega \propto k^2 \eeq
which gives transverse spin wave modes  which are gapless. The spin stiffness is proportional to superfluid density. Thus the density and spin are strongly coupled. Also the longitudinal spin mode cannot appear in the system since the length of the spins
are fixed in our model and thus such elastic vibrations are forbidden. These various collective modes will determine the nature of the stability of the quantum phases that we have described in our work.
One can also make some qualitative comments general nature of some collective oscillations above various ground state configurations. For example the collective oscillations over the spin current texture will intertwine the collective modes of vortices with higher winding number with spin waves and it will be very interesting to study these modes.

\section{Comparison with other methods \label{sec:compare}}
A natural question that arises how our method can be compared with the other methods,particularly described
in references \cite{Santos,diener} where the full quantum mechanical nature of the spin for Chromium BEC has been taken into account. A detailed comparison is  not possible due to different  treatment of the dipolar and short range interaction in both these works and  the type of order parameters that has been considered in reference \cite{diener}. However following issues can be pointed out. In eq. $3$ of reference \cite{Santos} the interaction strength of the short range interaction in various channels has been provided. It has been shown clearly there that all the spin dependent channels for the short range interaction is sub dominant as compared to rotationally invariant contact interaction in the maximum spin channel ($c_0$ in \cite{Santos}). Thus the basic assumption of our theoretical
formulation agrees with the analysis of this work. A major difference also comes from the single mode approximation adopted in that work, where as in our work the spin can rotate continuously as the field
$\theta$ and $\phi$ varies. This is why stabilization of spin current texture like state is more natural
in our method while there in \cite{Santos} one needs to go beyond SMA to observe such structures.
Additionally they have also assumed a polarized dipolar condensate as the equilibrium structure.
This is not the case in our analysis particularly for spin current texture as well as the hedgehog structure. Another difference in the methodology is that in the analysis of \cite{Santos} varying $\frac{g_0}{g_6}$ ( these are equivalent to interaction strengths in the singlet and triplet channel for spin-$1$ bosons) certain phases has been obtained. We have completely neglected the effect of singlet channel. Due to this the phases that can be stabilized by long wavelength fluctuations around the ferromagnetic phase can be observed in our analysis. However the phases which can be observed due to the condensation of spin-flip type of excitations around the ferromagnetic phases cannot be captured within our model. A possible extension of our model can be done by explicitly breaking the spin rotational symmetry that will allow such spin singlet excitations.

A comparison with the model proposed by Ho {\it et al.} is even more difficult since the order parameter considered there is very different. However given the general agreement between the phase diagram obtained in ref. \cite{Santos} and \cite{diener} the  differences between our  findings and those of reference \cite{diener} can be traced back to the same reasons we have just mentioned.

\section{SUMMARY AND OUTLOOK \label{sec:summary}}
In this paper we have reported the calculation of a a number of interesting equilibrium configuration
of the spinor dipolar Bose-Einstein condensates treating the dipole as a classical spin and neglecting
all other s-wave type interactions in various spin channels except the Hatree type repulsive one.
The resulting equilibrium configurations shows strong coupling between mass and spin density, indications
of quantum phase transitions and the presence of topological singularities in density and spin fields.
These spin textures can be observed directly via a novel phase-sensitive {\it in situ} detection \cite{review} or indirectly via conventional absorption imaging for the number density. Interesting magnetometric measurement is also possible to detect such spin-textures \cite{venga}.
For the future work, it is interesting to study vortex state, especially spin current texture. Since the density is already depleted for the spin current texture, such a vortex can be easily  observable and might become stable. We have  also  provided the theoretical framework in which the collective excitation spectrum
can be studied. In such excitations spin and density waves will get strongly coupled.
These are still difficult problem because of long-ranged anisotropic nature of the d-d interaction.
Moreover, the model Hamiltonian is applicable literally for electric dipolar systems without further approximation. We expect that BECs of hetero-nuclear molecules with permanent electric dipole moment might be realized in near future \cite{molecule} where the formation of textures presented here may be possible. Lastly, the above model being an effective theory it also has limitations. For example, the structures that can be attributed to the other spin-dependent channel of the s-wave interaction, cannot be obtained within this method. Including these effects in this description would be  highly interesting theoretical problem. Nevertheless the hallmark of this model is that it unfurls a number of topologically non trivial quantum phases and explain their existence in a very transparent method .
\section*{ACKNOWLEDGEMENTS}
%-----?t?I`?UN?UN?g??
We thank Tarun K. Ghosh and W. Pogosov for useful discussions in the early stage of this research. This work was supported by a grant of the Japan Society for the Promotion of Science. One of us (SG) thanks IIT Delhi
to provide financial support to attend the Grenoble BEC $2008$ conference where this work has been presented as a poster.
%----/

\end{document}